\documentclass[aps,pra,twocolumn,superscriptaddress]{revtex4}%
\usepackage{graphics}
\usepackage{amsmath}
\usepackage{graphicx}
\usepackage{amsfonts}
\usepackage{amssymb}
\usepackage{float}
\usepackage{longtable}
\usepackage{epsfig}
\usepackage{latexsym}
\usepackage{theorem}
\usepackage{bbm}
\usepackage{bm}
\usepackage{psfrag}

\begin{document}
\title{Detecting and tracking bacteria with quantum light}
\author{Gaetana Spedalieri}
\affiliation{Computer Science, University of York, York YO10 5GH, UK}
\affiliation{Research Laboratory of Electronics, Massachusetts Institute of Technology,
Cambridge MA 02139, USA}
\author{Lolita Piersimoni}
\affiliation{Department of Pharmaceutical Chemistry \& Bioanalytics, Institute of Pharmacy,
Martin-Luther University Halle-Wittenberg, Kurt-Mothes-Stra\ss e 3a, Halle
06120, Germany}
\affiliation{Biological Chemistry, Medical School, University of Michigan, Ann Arbor,
Michigan 48109, USA}
\author{Omar Laurino}
\affiliation{Smithsonian Astrophysical Observatory, 60 Garden Street, Cambridge, MA 02138, USA}
\author{Samuel L. Braunstein}
\affiliation{Computer Science, University of York, York YO10 5GH, UK}
\author{Stefano Pirandola}
\affiliation{Computer Science, University of York, York YO10 5GH, UK}

\begin{abstract}
The field of quantum sensing aims at improving the detection and estimation of
classical parameters that are encoded in physical systems by resorting to
quantum sources of light and quantum detection strategies. The same approach
can be used to improve the current classical measurements that are performed
on biological systems. Here we consider the scenario of two bacteria (\emph{E.
coli} and \emph{Salmonella}) growing in a Luria Bertani broth and monitored by
classical spectrophotometers. Their concentration can be related to the
optical transmissivity via the Beer-Lambert-Bouguer's law and their growth
curves can be described by means of Gompertz functions. Starting from
experimental data points, we extrapolate the growth curves of the two bacteria
and we study the theoretical performance that would be achieved with a quantum
setup. In particular, we discuss how the bacterial growth can in principle be
tracked by irradiating the samples with orders of magnitude fewer photons,
identifying the clear superiority of quantum light in the early stages of
growth. We then show the superiority and the limits of quantum resources in
two basic tasks: (i) the early detection of bacterial growth and (ii) the
early discrimination between two bacteria species.

\end{abstract}
\maketitle




\section{Introduction}

Growth curves are found in a wide range of disciplines, such as fishery
research, crop science, and other areas of biology~\cite{ref1}. They have for
a long time been used to study the dynamics of the populations of bacteria.
These curves typically show an initial lag time after which the concentration
(or number of organisms) starts to increase exponentially towards a maximal
saturation value; in this final stationary phase the growth rate gradually
decreases to zero as the asymptote is reached. This overall process results
into a typical sigmoidal curve that has been represented by various
mathematical models~\cite{ref3}. The parameters of this curve depend on the
specific process under study, be it bacterial growth in samples or
dose-mortality relations~\cite{ref2}.

In today's biology and chemistry laboratories, the spectrophotometer is the
instrument that is used to measure bacterial concentrations and therefore
track their growth. This is an optical instrument which measures the
transmission $\eta\in\lbrack0,1]$ of visible, UV or infrared light, through a
sample. More precisely, it measures the \textquotedblleft optical
absorbance\textquotedblright\ $A:=-\log_{10}\eta$~\cite{dB} also known as
\textquotedblleft optical density\textquotedblright\ (we assume negligible
scattering from the sample). Its basic principle is the well-known
Beer-Lambert-Bouguer law~\cite{Bouguer,Lambert,Beer}, which relates the
optical absorbance of a sample to its concentration~\cite{Ingle88}. More
precisely, the absorbance $A$ at some specific wavelength $\lambda$ is equal
to the concentration of the sample $C$ (in units of $\mathrm{mol}%
/\mathrm{m}^{3}$) times the length $l$ of the optical path (in units of
$\mathrm{m}$) times the molar extinction coefficient $\varepsilon$ specific of
the substance (in units of $\mathrm{m}^{2}/\mathrm{mol}$). Thus, we have the
formula $A=\varepsilon lC$ or equivalently $\eta=10^{-\varepsilon lC}$. In a
standard setup, where $\varepsilon$ and $l$\ are fixed, the absorbance is the
relevant quantity to be considered for tracking bacterial growth.


The two main types of spectrophotometers are single- and double-beam. The
first design measures only the light intensity at the output of the sample,
while the second measures the ratio of the light intensities at the output of
two separate paths, one sent through the sample and the other one sent through
a reference or blank.
It is important to estimate the typical number of photons that are irradiated
by these devices for their readout. To give an idea of the order of magnitude,
we perform a simple calculation based on one of the spectrophotometers that we
used for our experimental data (Ultraspect 2100 pro Amersham Bioscience). This
employs a Xenon light source at $600$nm with an average of power $10$W,
flashing at a frequency of $25$Hz, corresponding to about $0.4$J per flash. A
typical measurement involves about $6$ flashes for a total time of about $1/4$
of a second, corresponding to about $2.4$J of energy $E$ irradiated over the
sample. From Planck's law $E=nhc/\lambda$, we can derive the staggering value
of $n\simeq10^{19}$ thermally-distributed photons at $600$nm wavelength.

The irradiation of such a high energy is a potential disadvantage for this
instrument. In fact, a very high number of photons can be destructive,
especially if the sample contains photosensitive bacteria, fragile proteins or
DNA/RNA. Furthermore, there are other limitations to account for in current
spectrophotometers. One problem is the low sensitivity of the instrument,
which is often inadequate for good readouts of low concentrations, a task
which is very important in scenarios such as early disease detection or food
poisoning. Because of this poor performance, researchers may need to
re-prepare their samples many times to get a good statistical estimate.

In the present work, we discuss how the use of quantum resources can
drastically reduce the number of photons that are required for readouts of
bacterial concentration. Collecting experimental data with standard
spectrophotometers, first we study the typical realistic errors affecting
these classical instruments in tracking the growth of bacterial species
(\textit{E. coli} and \textit{Salmonella}). From this data, we extrapolate the
functional forms of the bacterial growth curves, which are then used in our
theoretical simulation of an optimal quantum setup. We show that similar
performances can be achieved by using quantum designs that employ sources of
light with orders of magnitude fewer photons, when suitably combined with
corresponding optimal quantum detections.

Depending on the working regime (lower or higher concentrations), there is a
preferable semiclassical or truly quantum state to be used for the input
light. At higher concentrations, one needs to consider coherent states
irradiating a relatively high mean number of photons per readout (e.g., of the
order of $10^{4}$). This source is studied in combination with an optimal
quantum detection at the output. It represents our semiclassical benchmark
which bounds the performance of any other classical source (even when the
output detection is quantum). We also discuss how its performance can be
achieved by using a double-beam setup where asymmetrically-correlated two-mode
thermal states are prepared at the input and photon-resolving measurements are
performed at the output.

Our results show that the use of truly quantum states is limited to low
concentrations, i.e., during the early phase of bacterial growth. Considering
this initial phase and assuming a small number of photons irradiated over the
sample, the performance of optimal quantum states in estimating the
concentration clearly outperforms the semiclassical benchmark based on
coherent states. In general, the optimal quantum states can be engineered as
suitable superpositions of number states~\cite{AdessoIlluminati,ReviewNEW} and
their quantum measurement is the optimal one which minimizes the quantum
Cramer Rao bound (QCRB)~\cite{SamMETRO}. In practice, we also analyze the
performance of quantum states (squeezed vacua) that can be easily engineered
in the lab, showing that they offer a similar quantum advantage as the optimal
quantum probes.

Because quantum advantage is relevant at low concentrations, it is therefore
important in tasks of early bacterial detection. We therefore study the task
of the early detection of the growth of \textit{E. coli }in a sample, and the
task of the early discrimination between the growth of two different bacterial
species (\textit{E. coli} and \textit{Salmonella}). In both cases, we are able
to show the advantage of the optimal and practical quantum sources with
respect to the semiclassical benchmark (coherent states), both in terms of
reducing the time for detecting bacterial growth and decreasing the error
probability in the discrimination between two different species.

The manuscript is organized as follows. In Sec.~\ref{Sec2} we describe our
classical experiments for tracking the growth of \textit{E. coli} and
\textit{Salmonella} via standard spectrophotometer in typical lab conditions.
Then, in Sec.~\ref{Sec3}, we study the performance of a theoretical
quantum-enhanced model of spectrophotometer based on semiclassical or truly
quantum sources and output quantum detection. In Sec.~\ref{Sec4} we study
early detection and discrimination of bacteria. Finally, Sec.~\ref{Sec5} is
for conclusions.

\section{Experimental growth curves with classical instruments\label{Sec2}}

We have performed two different experiments. In the first experiment, we
considered a single bacterial species (\textit{E. coli}, MRE600)~\cite{Meier}.
The results were averaged over the strain so as to consider an average
behavior. In the second experiment, we instead considered two different
species of bacteria (\textit{E. coli} BW25113) and \textit{Salmonella}
(\textit{enterica serovar Typhimurium} ATCC1428 strain) whose growth behaviors
were analyzed separately. In all cases, the bacteria were first grown in a
Luria Bertani broth at $30^{\circ}$C and then suitably diluted for subsequent
measurements. Their concentration (optical absorbance) were measured by using
classical spectrophotometers. Finally, the outcomes were statistically
post-processed into experimental growth curves. See Methods for more details.

From the experimental data, we extrapolated analytical forms for the growth
curves, according to the Gompertz function~\cite{Gompertz1825,ref3}. This
function relates the concentration/absorbance $A$\ of the sample to time $t$,
as follows%
\begin{equation}
A=a\exp[-\exp(b-ct)],
\end{equation}
where $a$, $b$, and $c$ are parameters to be interpolated from the data. Note
that the Gompertz function can also be re-written as~\cite[Eq.~(11)]{ref3}
\begin{equation}
A=a\exp\left\{  -\exp\left[  \frac{\mu e}{a}(\theta-t)+1\right]  \right\}
+A_{\text{bk}}, \label{GompertzF}%
\end{equation}
where $a$ is the asymptotic absorbance at infinite time $t\rightarrow\infty$,
$\mu:=ac/e$ is the rate of growth in the linear region, and $\theta:=(b-1)/c$.
Here we have also added an additional offset $A_{\text{bk}}$ accounting for
non-zero mean absorbance of the blank sample (i.e., non-unit transmissivity
$\eta_{\text{bk}}$ of the media holding the species under study).

\begin{figure}[ptbh]
\vspace{-0cm}
\par
\begin{center}
\includegraphics[width=0.39 \textwidth]{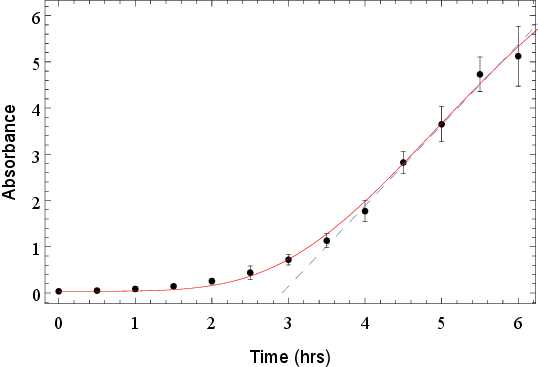} \vspace{-0.2cm}
\end{center}
\caption{Growth curve of \textit{E. coli} (wild type MRE 600) in terms of
optical absorbance versus time (hours). We show the experimental data,
suitably post-processed into a mean curve with error bars corresponding to one
standard deviation. The data is then interpolated by the Gompertz function
(red line) given by Eq.~(\ref{GompertzF}) with parameters $a\simeq9.4$,
$\mu\simeq1.7$, $\theta\simeq2.9$ and $A_{\text{bk}}\simeq0.036$. For
completeness, we also show the linear phase $(t-\theta)\mu$ of the growth
(dashed blue line). This linear phase occurs after the latency phase and
before the saturation phase of the sigmoid.}%
\label{Exp1Fig}%
\end{figure}

\begin{figure}[ptbh]
\vspace{+0.2cm}
\par
\begin{center}
\includegraphics[width=0.39 \textwidth]{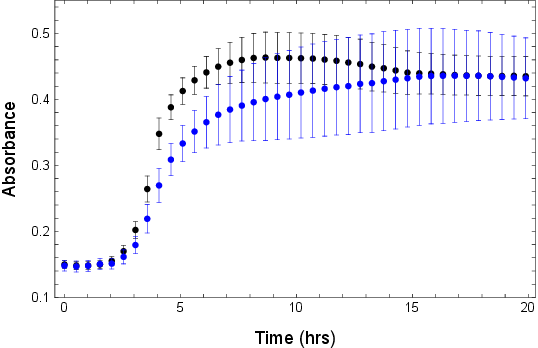} \vspace{-0.2cm}
\end{center}
\caption{Growth curves of \textit{E. coli} (black points)\ and
\textit{Salmonella} (blue points)\ in terms of optical absorbance versus time
(hours). We show the experimental data, post-processed into mean curves with
error bars corresponding to one standard deviation.}%
\label{Exp2Fig}%
\end{figure}

\begin{figure}[ptbh]
\vspace{-0.1cm}
\par
\begin{center}
\includegraphics[width=0.40 \textwidth]{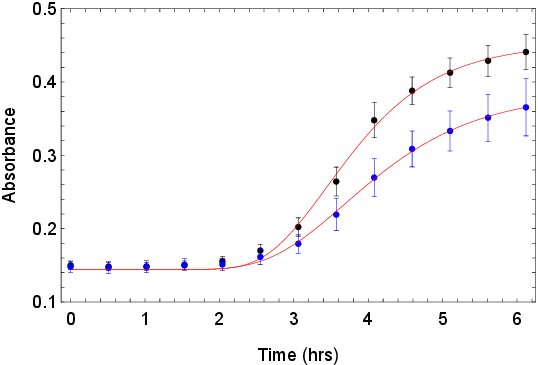} \vspace{-0.2cm}
\end{center}
\caption{Growth curves of \textit{E. coli} (black points) and
\textit{Salmonella} (blue points) in terms of optical absorbance versus time
(hours). We zoom the experimental data of Fig.~\ref{Exp2Fig} for times up to
six hours. The two sets of data points are interpolated by two Gompertz
functions with parameters {}$\{a,\mu,\theta\}\simeq\{0.309,0.139,2.634\}$ for
\textit{E. coli} and $\{a,\mu,\theta\}\simeq\{0.242,0.0882,2.672\}$ for
\textit{Salmonella}. In both cases, the blank has mean absorbance
$A_{\text{bk}}\simeq0.144$ (corresponding to a blank transmissivity
$\eta_{\text{bk}}\simeq0.717$).}%
\label{Exp2zoomFig}%
\end{figure}

The data of the first experiment is shown in Fig.~\ref{Exp1Fig}. At each time,
24 data points were measured and post-processed into a mean value plus an
error bar. Data was then used to interpolate a Gompertz function with suitable
parameters. The entire data of the second experiment is shown in
Fig.~\ref{Exp2Fig}. At each time, 18 data points per species were measured and
post-processed as before. In particular, in Fig.~\ref{Exp2zoomFig} we zoom on
the first six hours, where the two growth curves for \textit{E. coli} and
\textit{Salmonella} are more distinguishable. These experimental curves have
been interpolated by two Gompertz functions. 

\section{Theoretical performance with a quantum setup\label{Sec3}}

We now consider the theoretical performance that is achievable by using a
semiclassical or a quantum source at the input, combined with optimal quantum
detection at the output. Our first aim is to show that a semiclassical or
fully quantum setup can achieve an accuracy that is comparable with the
typical performance of a classical spectrophotometer while using orders of
magnitude fewer photons. As semiclassical sources, we consider single-mode
coherent states and also their approximation by means of two-mode correlated
thermal states. As truly quantum sources we consider optimal single-mode
states, such as number states and their superpositions~\cite{AdessoIlluminati}%
. To explore the limits achievable by these sources in the estimation of
optical absorbance, we consider the QCRB~\cite{SamMETRO}. For a fixed source
(input state) and number $N$ of probings of the sample, this bound provides
the minimum error-variance that we could achieve by optimizing over all
possible quantum measurements at the output.

First of all, for our purposes, we need to connect the error (standard
deviation) $\sigma_{\eta}$ affecting the transmissivity $\eta$ of the sample
to the error (standard deviation) $\sigma_{A}$\ associated with the absorbance
$A=-\log_{10}\eta$. A simple calculation provides $\sigma_{A}\simeq
\sigma_{\eta}/(\bar{\eta}\ln10)$, where $\bar{\eta}$ is the mean value of the
transmissivity, corresponding to $\bar{\eta}=10^{-\bar{A}}$, where $\bar{A}$
is the mean value of the absorbance. This approximation is justified by the
so-called delta-method~\cite{DM1,DM1b,DM2} (see Methods for more details). In
our theoretical simulation for the quantum setup, we assume that the
experimental mean value $\bar{A}$, which is well-approximated by the Gompertz
function, represents the actual physical value $A$\ of the absorbance.
Correspondingly, we assume that the mean value $\bar{\eta}=10^{-\bar{A}}%
$\ corresponds to the actual physical value of the transmissivity. As a
result, we may modify the previous expansion into the following form
\begin{equation}
\sigma_{A}\simeq\frac{\sigma_{\eta}}{\eta\ln10}. \label{scc}%
\end{equation}

The next step is to assume the QCRB for the computation of $\sigma_{\eta}$.
Assume that the sample can be approximately modelled as a pure-loss bosonic
channel $\mathcal{E}_{\eta}$ with transmissivity $\eta$. This channel/sample
is probed by $N$ quantum states $\rho_{\bar{n}}^{\otimes N}$ which irradiate a
total of $\bar{n}_{\text{tot}}:=N\bar{n}$ mean number of photons, where
$\bar{n}$ is the mean number of photons per state. Assuming an optimal
measurement of the output states $\rho_{\text{out}}^{\otimes N}$, we can
construct an unbiased estimator $\hat{\eta}$ of $\eta$. This estimator is
subject to an error-variance given by the QCRB%
\begin{equation}
\sigma_{\eta}^{2}\geq\frac{1}{NH_{\eta,\bar{n}}}, \label{QCRB}%
\end{equation}
where $H_{\eta,\bar{n}}$ is the quantum Fisher information
(QFI)~\cite{SamMETRO}, to be computed on the single-copy output state
$\rho_{\text{out}}:=\mathcal{E}_{\eta}(\rho_{\bar{n}})$. When $\rho
_{\text{out}}$ is a Gaussian state~\cite{RMP}, we can easily compute the QFI
from the fidelity, according to the general formulae in Ref.~\cite{Banchi}.

Combining Eqs.~(\ref{scc}) and~(\ref{QCRB}), we may write the following
standard deviation error for the absorbance%
\begin{equation}
\sigma_{A}\gtrsim\left[  \frac{1}{\eta(\ln10)\sqrt{NH_{\eta,\bar{n}}}}\right]
_{\eta=10^{-A}}. \label{main}%
\end{equation}
The explicit expression of the QFI\ $H_{\eta,\bar{n}}$ depends on the
transmissivity $\eta$ and the single-copy input state $\rho_{\bar{n}}$.
Assuming a single-beam configuration where the light emitted by the source can
be described by a coherent state $\rho_{\bar{n}}=\left\vert \sqrt{\bar{n}%
}\right\rangle \left\langle \sqrt{\bar{n}}\right\vert $ irradiating $\bar{n}%
$\ mean photons, we have $H_{\eta,\bar{n}}=\bar{n}/\eta$~\cite{MonrasParis},
so that
\begin{equation}
\sigma_{A}\gtrsim\frac{1}{\ln10}\sqrt{\frac{10^{A}}{\bar{n}_{\text{tot}}}}.
\label{scc2}%
\end{equation}
This performance can equivalently be achieved in a double-beam configuration
where a two-mode correlated thermal state is prepared in a very asymmetric
way, so that one mode is faint and the other is highly energetic (see
Ref.~\cite{Sped} for more details on this equivalence). The faint mode is sent
through the sample while the energetic one is directly sent to the output
measurement, where both the output modes are subject to photon counting (see
Methods for more details).

The optimal quantum performance corresponds to~\cite{MonrasParis}
$H_{\eta,\bar{n}}=\bar{n}[\eta(1-\eta)]^{-1}$, which is reached by input
number states or suitable superpositions~\cite{AdessoIlluminati}. By
substitution into Eq.~(\ref{main}), we derive the following improved error and
its expansion at low absorbance%
\begin{align}
\sigma_{A}  &  \gtrsim\frac{1}{\ln10}\sqrt{\frac{10^{A}-1}{\bar{n}%
_{\text{tot}}}}\label{optimalPP}\\
&  \simeq\sqrt{\frac{A}{\bar{n}_{\text{tot}}\ln10}}+\mathcal{O}(A^{3/2}).
\end{align}

An important observation about the standard deviations in Eqs.~(\ref{scc2})
and~(\ref{optimalPP}) is the fact that they depend on the energy of the input
via the mean total number of photons $\bar{n}_{\text{tot}}=N\bar{n}$. This
means that these quantities do not change if we consider a single energetic
state ($N=1$, $\bar{n}=\bar{n}_{\text{tot}}$) or a large number of
lower-energy states so that $N\gg1$ and $\bar{n}\ll\bar{n}_{\text{tot}}$ with
$N\bar{n}=\bar{n}_{\text{tot}}$ (assuming that the total measurement time of
this second option is reasonable). This is particularly useful for the truly
quantum resources which are typically limited to $1$ photon or less; for
these, we implicitly assume the condition of low-energy single-copy states so that we
increase the total energy by increasing the number of copies
$N$. Furthermore, in the regime of large $N$, the QCRB\ is known to be
achievable~\cite{SamMETRO,Paris,ReviewNEW} and the optimal detection strategy
can be realized by using local quantum measurements (i.e., performed over the single copies)
combined with adaptive estimators~\cite{asy1,asy2}.

\begin{figure}[ptbh]
\vspace{0.1cm}
\par
\begin{center}
\includegraphics[width=0.4 \textwidth]{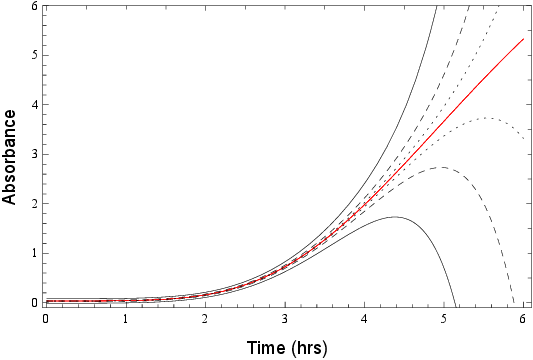} \vspace{-0.3cm}
\end{center}
\caption{Theoretical growth curves achievable by using coherent states and an
optimal output detection, reaching the QCRB in Eq.~(\ref{scc2}). We plot the
mean bacterial growth (solid red curve) as given by the Gompertz function from
Fig.~\ref{Exp1Fig}. Then, we consider the error bars (at one standard
deviation) given by coherent states with $\bar{n}_{\text{tot}}=100$ (solid
black lines), $\bar{n}_{\text{tot}}=1000$ (dashed black lines) and $\bar
{n}_{\text{tot}}=10000$ (dotted black lines).}%
\label{coherentPIC2}%
\end{figure}

\begin{figure}[ptbh]
\vspace{0.2cm}
\par
\begin{center}
\includegraphics[width=0.36 \textwidth]{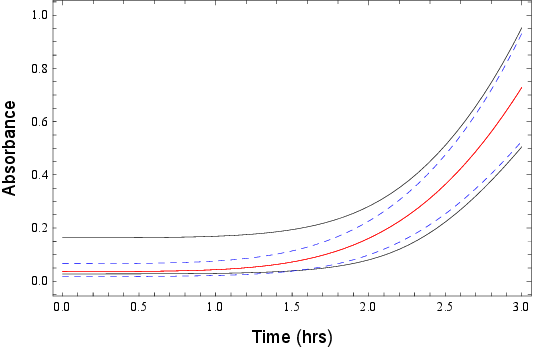} \vspace{-0.2cm}
\end{center}
\caption{Theoretical growth curves at low values of absorbance (latency
phase). By fixing $\bar{n}_{\text{tot}}=20$ photons, we compare the optimal
error bars achievable by coherent states (solid black lines) and those
achievable by the optimal quantum states (dashed blue lines). The mean growth
curve (solid red line) is the Gompertz function from Fig.~\ref{Exp1Fig}.}%
\label{numberPIC}%
\end{figure}

We show our numerical results in Figs.~\ref{coherentPIC2}-\ref{numberPIC},
considering the mean growth curve of \textit{E. coli} approximated by the
Gompertz function plotted in Fig.~\ref{Exp1Fig}. In Fig.~\ref{coherentPIC2} we
show the error bars (at one standard deviation) that we would obtain by using
coherent states for different values of total energy irradiated. As we can see
from the figure, the error bars are narrow at low absorbances for which we can
use relatively few photons, while they quickly increase at higher values of
the absorbance, for which we need to consider energetic coherent states. At
high absorbance, the performance of coherent states approximates the quantum
limit, as we can see by comparing Eqs.~(\ref{scc2}) and~(\ref{optimalPP}) for
large $A$. This means that, for this regime, it makes little sense to consider
truly quantum states such as number states and the best strategy is to use
coherent states with high energy.

However, different is the case for low absorbances/concentrations. As we can
see from Fig.~\ref{numberPIC}, at the early stage of bacterial growth, i.e.,
during the latency phase of the sigmoid, the use of optimal quantum sources
gives a non-trivial advantage with respect to coherent states, for the same
mean number of photons irradiated over the sample. In other words, the initial
latency phase, i.e., the low-concentration regime, is the most interesting
from the quantum point of view. Note the asymmetric behavior of the error bars
when the absorbance is close to zero. This is due to the fact that the
Gaussian distribution needs to be truncated. Start with a Gaussian
distribution with mean value $\bar{A}$ and standard deviation $\sigma_{A}$,
and imposes a $1$-sided truncation at the origin. Then the mean value and
standard deviation of the new distribution are given by%
\begin{equation}
\bar{A}^{\prime}=\bar{A}+g(\omega)\sigma_{A},~~\sigma_{A}^{\prime}=\sigma
_{A}\sqrt{1+\omega g(\omega)-g(\omega)^{2}},
\end{equation}
where $\omega:=-\bar{A}/\sigma_{A}$ and
\begin{equation}
g(x):=\frac{2\mathcal{N}(x)}{1-\operatorname{erf}\left(  x/\sqrt{2}\right)  },
\end{equation}
with $\mathcal{N}(x)$ being the standard normal distribution and
$\operatorname{erf}(x):=2\pi^{-1/2}\int_{0}^{x}e^{-x^{2}}dx$ the error
function. 

Besides investigating the optimal quantum limit, in our following analysis we
also explore the performance of a practical quantum source, as represented by
an input squeezed vacuum state. This is a zero-mean Gaussian state with
covariance matrix $\mathrm{diag}(r,r^{-1})$, where $r\leq1$ represents the
squeezing parameter (in the $q$-quadrature). It is easy to see that its mean
number of photons depends on $r$ and is given by $\bar{n}=(1-r)^{2}/(4r)$.
From the literature~\cite{ReviewNEW,MonrasParis}, we know that the performance
of this input state for bosonic loss estimation is close to optimal when
$\bar{n}$ is low. In particular, adapting the result from
Ref.~\cite{MonrasParis}, the QFI takes the form%
\begin{equation}
H_{\eta,\bar{n}}=\frac{1}{\eta}\left[  \frac{\bar{n}-2\bar{n}\eta(1-\eta
)}{(1-\eta)[1+2\bar{n}\eta(1-\eta)]}\right]  , \label{Hsqueeze}%
\end{equation}
which can be used in Eq.~(\ref{main}) to derive the performance of the
squeezed vacuum probe.

Note that, in this case, parameters $\bar{n}$ and $N$ are not simply combined
as before, but we need to fix $N$ and use $\bar{n}=\bar{n}_{\text{tot}}/N$ in
Eq.~(\ref{Hsqueeze}). In this case, we get a lower bound from Eq.~(\ref{main})
which depends on $A$ and $\bar{n}_{\text{tot}}$. More precisely, one may start
by fixing the amount of decibels that can be realized for the squeezed probes.
Even though values up to $9.3$~dB are currently
realizable~\cite{SqueezingTobias}, in our study we will consider the case of a
relatively cheap source with just $1$~dB of squeezing, so that $r=10^{-1/10}%
\simeq0.794$ and we have $\bar{n}\simeq0.013$ mean photons per probe. For
instance, this means that using a total of $\bar{n}_{\text{tot}}=20$ mean
photons corresponds to irradiating $N=1500$ probes. Number of probes in the
order of $10^{3}-10^{5}$\ requires a negligible time with respect to the
time-scale associated with the bacterial growth. For instance, using a clock
rate and a detector at $1$ MHz, these probes are generated and detected in
about $1-100$~ms.



\section{Quantum-enhanced early detection\label{Sec4}}

\subsection{Detecting growth of \textit{E. coli}}

Once it is understood that the initial phase of the bacterial growth is the
most interesting one from the quantum point of view, we therefore consider the
task of early detection. This consists in distinguishing whether bacteria are
growing or not in the sample. More precisely, we study the time required for
successfully discriminating whether the sample is blank or contains \textit{E.
coli} growing in accordance with the experimental data of our first experiment
(see Fig.~\ref{Exp1Fig}). As a first step, we translate the absorbance data
$A$ into transmissivity data $\eta=10^{-A}$ which is the quantity physically
measured by the instrument (and following a Gaussian distribution under the
assumption of many measurements). During the first phase of the growth (up to
$3$ hours), we interpolate the experimental data with a theoretical curve of
the form $\eta(t)=\eta_{\text{bk}}-ct^{2}+dt^{3}$, where $\eta_{\text{bk}}$ is
the transmissivity of the blank sample, while $c$ and $d$ are suitable
constants. From the experimental data of the absorbance (see
Fig.~\ref{Exp1Fig}), we therefore retrieve the corresponding decay in optical
transmissivity $\eta$ versus time $t$ (hours) for \textit{E. coli}. We find
$\eta_{\text{bk}}\simeq0.92$, $c\simeq0.1$ and $d\simeq0.0088$ for the cubic
theoretical curve $\eta(t)$.

\begin{figure}[ptbh]
\vspace{0.2cm}
\par
\begin{center}
\includegraphics[width=0.36 \textwidth]{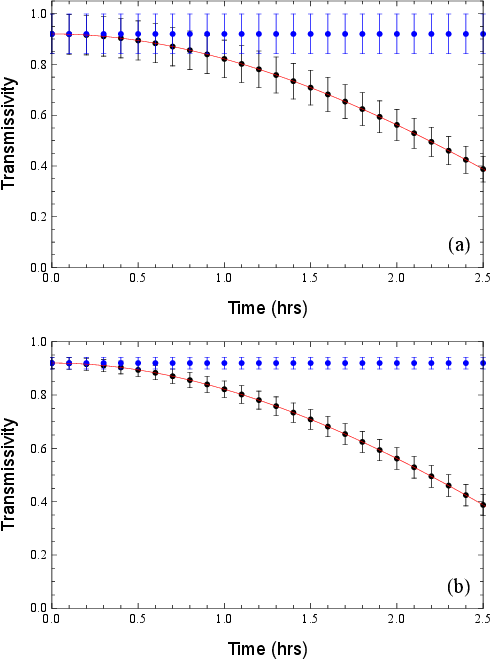} \vspace{-0.2cm}
\end{center}
\caption{Decay of transmissivity due to the growth of \textit{E.\ coli} versus
time $t$\ (black points) compared with the constant transmissivity of a blank
sample (blue points). The sample is monitored by irradiating a total of
$\bar{n}_{\text{tot}}=150$ photons for each reading. Error bars refer to one
standard deviation as given by the QCRB. In (a)~we consider coherent states as
the input source, while in (b) we consider optimal quantum states. The mean
decay (red line) is given by the curve $\eta(t)$ extrapolated by the
experimental data and described in the main text.}%
\label{compEXP1PIC}%
\end{figure}

Using the curve $\eta(t)$ we then consider the error bars achievable by an
optimal quantum setup (in terms of source and detection) and those that are
instead achievable by a semi-classical setup where the source is prepared in
coherent states and the output is optimally detected by a quantum measurement.
As previously discussed, the latter is a benchmark which bounds the ultimate
theoretical performance of any classical setup, i.e., based on classical
sources (coherent/thermal states) combined with classical receivers (e.g.,
non-photon-resolving intensity measurements).

Using the QCRB for coherent states $\sigma_{\eta}\geq\sqrt{\eta/\bar
{n}_{\text{tot}}}$ and the QCRB for the optimal quantum states $\sigma_{\eta
}\geq\sqrt{\eta(1-\eta)/\bar{n}_{\text{tot}}}$, we plot the curves in
Fig.~\ref{compEXP1PIC}. This figure already qualitatively shows that truly
quantum sources can perform much better at short times. Below we make this
observation quantitative by computing the corresponding error probabilities in
detecting the bacterial growth as a function of time.

Let us assume that at time $t$, we can perform a sufficiently large number of
measurements, so that the QCRB is well-approximated (we use many probes $N$,
each with small mean number of photons $\bar{n}$ such that the total $N\bar
{n}$ matches the fixed energetic constraint $\bar{n}_{\text{tot}}$). At each
reading time $t$, the data points $\{\eta_{k}\}_{k=1}^{N}$ provided by the
quantum measurement are used to build an estimator $\hat{\eta}$ of the
transmissivity $\eta(t)$ whose error $\sigma_{\eta}$ is given by the QCRB for
$\bar{n}_{\text{tot}}$ mean total number of photons irradiated by the source.
Assume that the estimator approximately follows a Gaussian distribution in
$\eta$ as a result of the central limit theorem (e.g., the estimator may be
based on the arithmetic mean of the outcomes which, in turn, are identically
and independently distributed). Furthermore, for increasing $\bar
{n}_{\text{tot}}$, the standard deviation $\sigma_{\eta}$ of this distribution
is sufficiently small, so that the truncation of the tails at the border of
the finite segment $0\leq\eta\leq1$ becomes a relatively small effect.

For the null hypothesis $H_{0}$ (no growth), the estimator $\hat{\eta}$ is
centered around $\eta_{\text{bk}}$ according to a Gaussian distribution
$p_{0}(\eta)$ with standard deviation $\sigma_{\eta_{\text{bk}}}$. For the
alternative hypothesis $H_{1}$ (yes growth), the estimator $\hat{\eta}$ will
be instead centered around $\eta(t)\leq\eta_{\text{bk}}$ according to a
Gaussian distribution $p_{1}(\eta)$ with standard deviation $\sigma_{\eta(t)}%
$. We can therefore consider a decision test with threshold $0\leq\tau\leq1$:
if $\hat{\eta}\geq\tau$ we accept the null hypothesis $H_{0}$, while if
$\hat{\eta}<\tau$ we accept the alternative hypothesis $H_{1}$. Consequently,
there are associated false-positive $p_{\text{FP}}$ and false-negative
$p_{\text{FN}}$\ error probabilities%
\begin{align}
p_{\text{FP}}  &  :=\text{prob}(H_{1}|H_{0})=\mathcal{N}_{0}^{-1}\int
_{0}^{\tau}p_{0}(\eta)d\eta,\label{FP1}\\
p_{\text{FN}}  &  :=\text{prob}(H_{0}|H_{1})=\mathcal{N}_{1}^{-1}\int_{\tau
}^{1}p_{1}(\eta)d\eta, \label{FN1}%
\end{align}
where the normalization factors $\mathcal{N}_{i}:=\int_{0}^{1}p_{i}(\eta
)d\eta$ for $i=0,1$ are due to the truncation at the border. Under this
hypothesis, we may compute
\begin{align}
p_{\text{FP}}(\tau)  &  =\frac{1}{2\mathcal{N}_{0}}\left\{  \operatorname{erf}%
\left[  \frac{\eta_{\text{bk}}}{\sqrt{2}\sigma_{\eta_{\text{bk}}}}\right]
-\operatorname{erf}\left[  \frac{\eta_{\text{bk}}-\tau}{\sqrt{2}\sigma
_{\eta_{\text{bk}}}}\right]  \right\}  ,\\
p_{\text{FN}}(\tau,t)  &  =\frac{1}{2\mathcal{N}_{1}}\left\{
\operatorname{erf}\left[  \frac{\eta(t)-\tau}{\sqrt{2}\sigma_{\eta(t)}%
}\right]  -\operatorname{erf}\left[  \frac{\eta(t)-1}{\sqrt{2}\sigma_{\eta
(t)}}\right]  \right\}  .
\end{align}
\begin{figure}[ptbh]
\vspace{0.2cm}
\par
\begin{center}
\includegraphics[width=0.36 \textwidth]{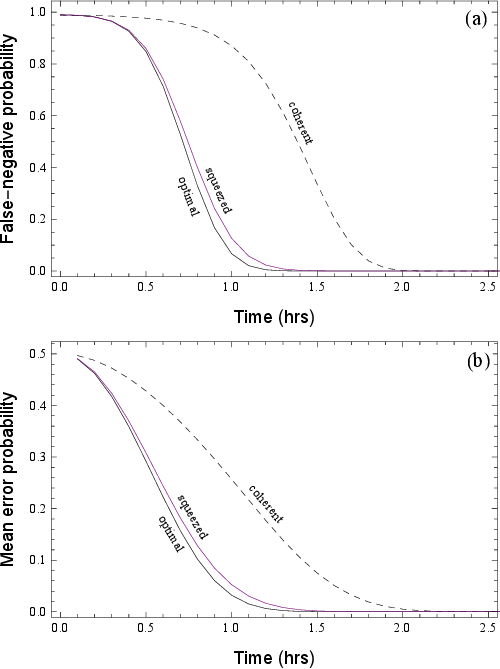} \vspace{-0.2cm}
\end{center}
\caption{{}Early detection of growth of \textit{E. coli}. We plot the error
probability versus time $t$ (hours) for the semiclassical case of a
coherent-state source (dashed line), an optimal quantum source (solid black
line) and a practical squeezed quantum source (solid purple line), all of them
irradiating $\bar{n}_{\text{tot}}=150$ mean total photons over the sample at
each reading time. In particular, we choose $1$dB of squeezing for the
squeezed vacuum probe, so that we have $\bar{n}\simeq1.33\times10^{-2}$ mean
photons per mode, which requires the use of $N=11267$ probes (so as to
irradiate $150$ mean photons overall). In panel (a) we consider the
false-negative error probability $p_{\text{FN}}$ over time $t$, fixing the
value of false-positive error probability to $p_{\text{FN}}=1\%$. In panel (b)
we plot the mean error probability $p_{\text{mean}}$ over time $t$. For both
symmetric and asymmetric testing, we can see how the optimal and practical
quantum sources allow one to detect bacterial growth much earlier than the
semiclassical benchmark (at about $1$ hour instead of $2$ hours).}%
\label{bothTESTSpic}%
\end{figure}

We have now two possible types of testing. In asymmetric testing, we fix a
tolerable value for the false positives. This means we fix a value for
$p_{\text{FP}}$ and, therefore, for the threshold parameter $\tau$, which can
be expressed as an inverse function $\tau=\tau(p_{\text{FP}})$. We then
replace $\tau$ in $p_{\text{FN}}(\tau,t)$, and study the false-negative error
probability $p_{\text{FN}}$ over time. In symmetric testing, we instead assume
that the two error probabilities have equal Bayesian costs. In the case of the
same priors, the quantity of interest is therefore the mean error probability
\begin{equation}
p_{\text{mean}}(t):=\min_{\tau}\frac{p_{\text{FP}}(\tau)+p_{\text{FN}}%
(\tau,t)}{2}. \label{PmeanEQ}%
\end{equation}
The numerical results are shown in Fig.~\ref{bothTESTSpic} for both asymmetric
and symmetric testing. In the regime of small energy ($\bar{n}_{\text{tot}%
}=150$ in the figure), we can see that optimal quantum states allow us to
detect the growth of \textit{E. coli} about$\ 1$ hour earlier than coherent
states. Approximately the same quantum advantage can be reached by using as
input source of light composed of a tensor-product of squeezed vacuum states
with just $1$~dB of squeezing.

\subsection{Discrimination of different bacterial species}

\begin{figure}[ptbh]
\vspace{0.2cm}
\par
\begin{center}
\includegraphics[width=0.36 \textwidth]{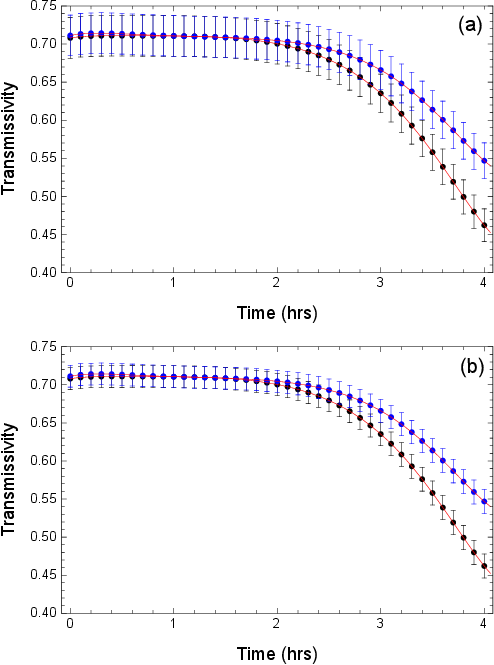} \vspace{-0.2cm}
\end{center}
\caption{Decay of transmissivity due to the growth of \textit{E.\ coli} (black
points) and Salmonella (blue points) versus time $t$ (hours). The sample is
monitored by irradiating a total of $\bar{n}_{\text{tot}}=10^{3}$ mean photons
for each reading. Error bars refer to one standard deviation as given by the
QCRB. In (a)~we consider coherent states as the input source, while in (b) we
consider optimal quantum states. The mean decay (red lines) are given by
curves $\eta_{\text{Ecoli}}(t)$ and $\eta_{\text{Salmo}}(t)$ that are
extrapolated by the experimental data from Fig.~\ref{Exp2zoomFig}.}%
\label{transECSPIC}%
\end{figure}

To further explore this capability, let us also study the performance in the
early discrimination between different bacteria, starting from the
experimental data obtained for \textit{E. coli} and \textit{Salmonella} (see
Fig.~\ref{Exp2zoomFig}). As before the experimental data in absorbance $A$ can
be expressed in terms of the transmissivity $\eta=10^{-A}$ and the
corresponding growth curves of the two bacteria can be interpolated by two
polynomial functions $\eta_{\text{Ecoli}}(t)$ and $\eta_{\text{Salmo}}(t)$. At
each reading time $t$, the data points $\{\eta_{k}\}_{k=1}^{N}$ of a
theoretical quantum measurement provide an estimator $\hat{\eta}$ of the
transmissivity $\eta(t)$. The minimum error $\sigma_{\eta}$ will be given by
the QCRB relative to the specific source and the mean total number of photons
$\bar{n}_{\text{tot}}$ irradiated over the sample. The numerical performances
of coherent states and optimal quantum states are shown in
Fig.~\ref{transECSPIC}, up to $4$ hours. We can see that, while the quantum
source certainly narrows the error bars, the early discrimination between the
two bacteria appear to be more difficult than detecting a generic growth with
respect to the blank.

\begin{figure}[ptbh]
\vspace{0.2cm}
\par
\begin{center}
\includegraphics[width=0.36 \textwidth]{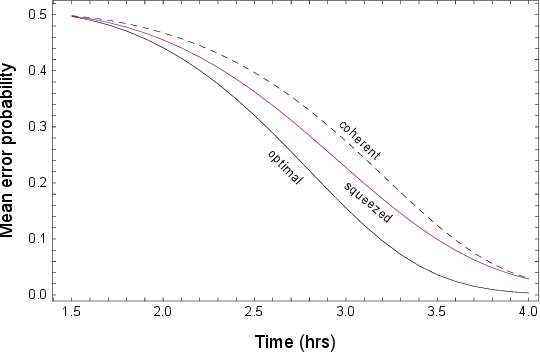} \vspace{-0.2cm}
\end{center}
\caption{Early symmetric discrimination of \textit{E. coli} and
\textit{Salmonella}. We plot the mean error probability versus time $t$
(hours) for a coherent-state source (dashed line), an optimal quantum source
(solid black line), and a practical quantum source (solid purple line), all
irradiating $\bar{n}_{\text{tot}}=10^{3}$ mean total photons per reading. For
the practical quantum source, we consider a tensor-product of $1$~dB squeezed
vacua, so that we need $N=75113$ probes per reading.}%
\label{bacteriapic}%
\end{figure}

For the null hypothesis $H_{0}$ (growth of \textit{Salmonella}), the estimator
$\hat{\eta}$ is centered around $\eta_{\text{Salmo}}(t)$ according to a
Gaussian distribution $p_{0}(\eta)$ with standard deviation $\sigma
_{\eta_{\text{Salmo}}(t)}$. For the alternative hypothesis $H_{1}$ (growth of
\textit{E. coli}), the estimator $\hat{\eta}$ will be instead centered around
$\eta_{\text{Ecoli}}(t)$ according to a Gaussian distribution $p_{1}(\eta)$
with standard deviation $\sigma_{\eta_{\text{Ecoli}}(t)}$. As before, we
consider a decision test with threshold $0\leq\tau\leq1$: if $\hat{\eta}%
\geq\tau$ we accept the null hypothesis $H_{0}$, while if $\hat{\eta}<\tau$ we
accept the alternative hypothesis $H_{1}$. The associated false-positive
$p_{\text{FP}}$ and false-negative $p_{\text{FN}}$\ error probabilities are
defined as in Eqs.~(\ref{FP1}) and~(\ref{FN1}). From these probabilities
$p_{\text{FP}}(\tau,t)$ and $p_{\text{FN}}(\tau,t)$, we can construct the mean
error probability $p_{\text{mean}}(t):=\min_{\tau}[p_{\text{FP}}%
(\tau,t)+p_{\text{FN}}(\tau,t)]/2$ for equal priors. We compare this mean
error probability assuming coherent state sources, optimal and practical
quantum sources irradiating the same mean number of total photons $\bar
{n}_{\text{tot}}$ per reading. As depicted in Fig.~\ref{bacteriapic}, an
optimal quantum source gives a clear advantage in the early discrimination
between the two bacteria, even though the advantage seems to be reduced to
less than one hour (about 30 minutes). This quantum advantage is further
reduced but yet present when considering the practical quantum source (based
on $1$~dB squeezed vacuum). 

\section{Conclusion\label{Sec5}}

In this work, we have explored the potentialities of a quantum-enhanced model
of spectrophotometer in detecting and tracking bacterial growth in samples.
Starting from experimental growth curves of two bacteria, \textit{E.~coli} and
\textit{Salmonella}, we simulate the theoretical performance achievable by a
quantum design that is based on an input source, semiclassical or a
truly-quantum, combined with an optimal quantum measurement at the output. We
first discuss how this device could efficiently work with orders of magnitude
fewer photons, also identifying the regime (low concentrations/absorbances)
where quantum sources can provide a non-trivial advantage. We have further
explored this regime considering tasks of early detection of bacterial growth
and early discrimination between two bacterial species. In each case, we have
shown that truly-quantum light allows us to improve the
detection/discrimination performance with respect to the use of coherent
states.

In conclusion, our work contributes to clarify the potentialities of
noninvasive quantum sensing techniques for biological and biomedical
applications. Further investigations may be aimed at the exploration of
similar advantages for other types of bacteria, the explicit design of optimal
receivers, and the simultaneous discrimination of multiple ($>2$) species
growing within a sample.

\bigskip

\textbf{Acknowledgments.}~This work has been sponsored by\ European Union's
Horizon 2020 Research and Innovation Action under grant agreements No. 745727
(Marie Sk\l odowska-Curie Global Fellowship \textquotedblleft quantum sensing
for biology\textquotedblright, QSB)\ and No. 862644 (\textquotedblleft Quantum
readout techniques and technologies\textquotedblright, QUARTET). GS and LP
would like to thank Janine Maddock, Mattew Chapman and Janet Price for
discussions. OL acknowledges expertise gained while working at the Chandra
X-ray Center. OL contributed to this work in his own time, not as part of his
CXC duties.

\bigskip

\textbf{Author Contributions.}~GS and SP conceived the idea, developed the
scheme, and proved both the analytical results and the numerical results
displayed in the plots. GS and LP performed the experiments with the classical
instruments. GS and OL performed the data analysis of the experimental data.
SP and GS wrote the manuscripts with edits by SLB.

\section*{Methods}

\subsection*{Description of the experiments}

In the first experiment, we have averaged over a single strain of
\textit{E.~coli} MRE600~\cite{Meier}. The strain was measured while growing
in\ a Luria Bertani (LB) broth at $30^{\circ}$C. In particular, three
different colonies of the MRE600 strain were selected from a Petri plate and
incubated overnight. Each colony was then re-suspended in 5ml LB and let grow
at $30^{\circ}$C overnight. Subsequently, each culture has been diluted 1:100
in new flasks containing fresh LB (a total of 3 flasks), so that the initial
optical density (OD) at 600 nm was 0.02 for all of them. The new cultures were
incubated at $30^{\circ}$C and the OD was measured at various times with 4
different dilution (1:1; 1:2; 1:5 and 1:10) with a technical replicate for
each dilution for a total of $24$ samples. The duration of the experiment was
6~hours and the measurements were performed by using a single-beam
spectrophotometer (Ultraspect 2100 pro Amersham Bioscience). The results of
the readings were then post-processed~\cite{Data}. During
the post-processing analysis, some of the data points were filtered
considering the appropriate dilutions and the fact that readings of OD that
are greater than $1$ are not reliable. We call OD$_{d}$ the optical density
measured for a 1:$d$ diluted sample. We only accept measured values such that
OD$_{d}<1$. Then, we compute the effective (non-diluted) OD\ of the sample as
$d^{-1}$OD$_{d}$ which is the quantity plotted in Fig.~\ref{Exp1Fig}. At each
measurement time, the appropriate readings from all the strains were combined
to form a single vector of $24$ data points, over which we computed average
and standard deviation.

In the second experiment, we have analyzed a strain of \textit{E. coli}
BW25113~\cite{Meier} and a strain of \textit{Salmonella} (textit{enterica
serovar Typhimurium} ATCC 14028). LB broth was used again to grow the two
species at $30^{\circ}$C. As before, 3 colonies of each species were grown in
different test tubes overnight and later diluted (roughly 1:100) in new test
tubes with fresh LB in order to have all the cultures at the same starting
point (around OD of 0.02 at 600 nm). Each tube was then used to provide 6
samples for a total of 18 samples per species. These 18 samples were
transferred to the micro-plate of an automatic spectrophotometer (infinite
M200 Pro microplate reader by Tecan). This particular instrument performed
readings of the $18$ samples every $30$ minutes for $20$ hours. The
contribution of the blank was estimated from $4$ blank samples also measured
every $30$ minutes for $20$ hours, for a total of $4\times40=160$
measurements. The blank contribution to the absorbance was equal to
$0.144\pm0.006$. The results of the readings were then post-processed~\cite{Data}.

\subsection{Delta method}

In general, consider a sequence of random variables $X_{n}$ converging in
distribution to a normal variable $X$ with (finite) mean value $\bar{X}$ and
(finite) variance $\mathrm{var}(X)$. Convergence in distribution means that
the cumulative function $F_{n}$ of $X_{n}$ converges to the cumulative
function $F$ of $X$, pointwise in the entire region where $F$ is continuous.
Now take a differentiable function $A(X)$ with non-zero first derivative.
Then, the sequence $A(X_{n})$ converges in distribution to a limit variable,
which is normal with mean value $A(\bar{X})$ and variance $[A^{\prime}(\bar
{X})]^{2}\mathrm{var}(X)$. This is the case when $A(X)=-\log_{10}(X)$ for
which we have $[A^{\prime}(\bar{X})]^{2}=[(\mathrm{ln}10)\bar{X}]^{-2}$. For
sufficiently large number of probings, we can assume, with good approximation,
that the transmissivity $\eta$ is distributed normally around the mean value
$\bar{\eta}$ with small standard deviation $\sigma_{\eta}$. Therefore, we can
write the first-order approximation $\sigma_{A}\simeq\sigma_{\eta}/(\bar{\eta
}\ln10)$.

\subsection*{Performance of correlated-thermal states}

The formulas in the main text refer to single-mode sources. Let us here
consider a two-mode source, therefore suitable for a double-beam design. In
particular, we consider a two-mode correlated thermal state combined with a
practical quantum detection at the output based on photon counting.

Recall that a two-mode correlated thermal state is a zero-mean Gaussian state
with covariance matrix~\cite{RMP}%
\begin{equation}
\mathbf{V}_{AB}=\left(
\begin{array}
[c]{cc}%
a\mathbf{I} & c\mathbf{I}\\
c\mathbf{I} & b\mathbf{I}%
\end{array}
\right)  ,
\end{equation}
where $\mathbf{I}=\mathrm{diag}(1,1)$ and%
\begin{align}
a  &  :=\bar{n}+1/2,~b:=\bar{n}(x^{-1}-1)+1/2,\label{eqb}\\
c  &  :=\sqrt{(1-x)/x}\bar{n}.
\end{align}
Here $\bar{n}$ is the mean number of thermal photons in the mode $A$
irradiated over the sample, while $0<x<1$ is an asymmetry parameter, so that
mode $B$ contains $\bar{n}(x^{-1}-1)$ mean thermal photons. We perform photon
counting on the output modes $A$ (sent through the sample with transmissivity
$\eta$) and $B$ (directly sent to the receiver).

The optimal performance is given by the classical Cramer-Rao bound
\begin{equation}
\sigma_{\eta}^{2}\geq\frac{1}{Nh_{\eta,\bar{n},x}},
\end{equation}
where the classical Fisher information $h_{\eta,\bar{n},x}$ is~\cite{Sped}%
\begin{equation}
h_{\eta,\bar{n},x}=\frac{\gamma\bar{n}}{\eta},~\gamma:=\frac{1+(1-x)\bar
{n}x^{-1}}{1+(1-x+x\eta)\bar{n}x^{-1}}\leq1. \label{FIz}%
\end{equation}
Using Eq.~(\ref{scc}), we therefore find%
\begin{equation}
\sigma_{A}\gtrsim\frac{1}{\ln10}\sqrt{\frac{10^{A}}{\gamma\bar{n}_{\text{tot}%
}}}, \label{thBB}%
\end{equation}
where $\gamma=\gamma(x,\bar{n},A)$ by replacing $\eta=10^{-A}$ in
Eq.~(\ref{FIz}). For fixed absorbance $A$ and input energy $\bar{n}$, we can
optimize $\sigma_{A}$ over $x$. For large asymmetry $x\rightarrow+\infty$, we
get $\gamma\rightarrow1$, so that Eq.~(\ref{thBB}) becomes equal to
Eq.~(\ref{scc2}) which is the optimal performance achievable by coherent
states (with an optimal quantum measurement).


\begin{thebibliography}{99}
\bibitem {dB}Note that this definition is close to the quantification of loss
in decibels, which includes an additional factor $10$, i.e., $-10\log_{10}%
\eta$.
\bibitem {Data}Data and code for figures are available at the doi: 10.5281/zenodo.4127011.
\end{thebibliography}
\end{document}